\documentclass[prc,aps,amsfonts,amsmath,superscriptaddress, twocolumn,showpacs,floatfix, altaffilletter, nofootinbib]{revtex4-1}

\usepackage{amsfonts}
\usepackage{amsmath}
\usepackage{amssymb}
\usepackage{bm}
\usepackage{graphicx}
\usepackage{hyperref}
\hypersetup{
     colorlinks,
     linkcolor={red!50!black},
     citecolor={blue!50!black},
     urlcolor={blue!80!black}
}
\usepackage{textcomp}
\usepackage[table]{xcolor}

\usepackage{multirow}

\usepackage{array}
\newcolumntype{M}[1]{>{\centering\arraybackslash}m{#1}}

\begin{document}

\title{\bf Production cross section and decay study of $^{243}$Es and $^{249}$Md}

\author{R.~Briselet}
\affiliation{Irfu, CEA, Universit\'e Paris-Saclay, F-91191 Gif-sur-Yvette, France}
\author{Ch.~Theisen}
\email[E-mail: ]{christophe.theisen@cea.fr}
\affiliation{Irfu, CEA, Universit\'e Paris-Saclay, F-91191 Gif-sur-Yvette, France}
\author{M.~Vandebrouck}
\affiliation{Irfu, CEA, Universit\'e Paris-Saclay, F-91191 Gif-sur-Yvette, France}
\author{A.~Marchix}
\affiliation{Irfu, CEA, Universit\'e Paris-Saclay, F-91191 Gif-sur-Yvette, France}
\author{M.~Airiau}
\affiliation{Irfu, CEA, Universit\'e Paris-Saclay, F-91191 Gif-sur-Yvette, France}
\author{K.~Auranen}
\altaffiliation{Present address: Physics Division, Argonne National Laboratory, 9700 South Cass Avenue, Lemont, Illinois 60439, USA}
\affiliation{University of Jyvaskyla, Department of Physics, P.O. Box 35, FI-40014 Jyvaskyla, Finland}
\author{H.~Badran}
\affiliation{University of Jyvaskyla, Department of Physics, P.O. Box 35, FI-40014 Jyvaskyla, Finland}
\author{D.~Boilley}
\affiliation{Grand Acc\'el\'erateur National d'Ions Lourds (GANIL), CEA/DSM - CNRS/IN2P3, Bd Henri Becquerel, BP 55027, F-14076 Caen Cedex 5, France}
\affiliation{Normandie Universit\'e, UNICAEN, Caen, France}
\author{T.~Calverley}
\affiliation{University of Jyvaskyla, Department of Physics, P.O. Box 35, FI-40014 Jyvaskyla, Finland}
\affiliation{University of Liverpool, Department of Physics, Oliver Lodge Laboratory, Liverpool L69 7ZE, United Kingdom}
\author{D.~Cox}
\altaffiliation{Present address: University of Lund, Box 118, 221 00 Lund, Sweden}
\affiliation{University of Jyvaskyla, Department of Physics, P.O. Box 35, FI-40014 Jyvaskyla, Finland}
\affiliation{University of Liverpool, Department of Physics, Oliver Lodge Laboratory, Liverpool L69 7ZE, United Kingdom}
\author{F.~D\'echery}
\affiliation{Irfu, CEA, Universit\'e Paris-Saclay, F-91191 Gif-sur-Yvette, France}
\affiliation{Institut Pluridisciplinaire Hubert Curien, F-67037 Strasbourg, France}
\author{F.~Defranchi Bisso}
\affiliation{University of Jyvaskyla, Department of Physics, P.O. Box 35, FI-40014 Jyvaskyla, Finland}
\author{A.~Drouart}
\affiliation{Irfu, CEA, Universit\'e Paris-Saclay, F-91191 Gif-sur-Yvette, France}
\author{B.~Gall}
\affiliation{Institut Pluridisciplinaire Hubert Curien, F-67037 Strasbourg, France}
\author{T.~Goigoux}
\affiliation{Irfu, CEA, Universit\'e Paris-Saclay, F-91191 Gif-sur-Yvette, France}
\author{T.~Grahn}
\affiliation{University of Jyvaskyla, Department of Physics, P.O. Box 35, FI-40014 Jyvaskyla, Finland}
\author{P.~T.~Greenlees}
\affiliation{University of Jyvaskyla, Department of Physics, P.O. Box 35, FI-40014 Jyvaskyla, Finland}
\author{K.~Hauschild}
\affiliation{CSNSM, IN2P3-CNRS, F-91405 Orsay Campus, France}
\author{A.~Herzan}
\altaffiliation{Present address: Institute of Physics, Slovak Academy of Sciences, SK-84511 Bratislava, Slovakia}
\affiliation{University of Jyvaskyla, Department of Physics, P.O. Box 35, FI-40014 Jyvaskyla, Finland}
\author{R.~D.~Herzberg}
\affiliation{University of Liverpool, Department of Physics, Oliver Lodge Laboratory, Liverpool L69 7ZE, United Kingdom}
\author{U.~Jakobsson}
\altaffiliation{Present address: Laboratory of Radiochemistry, Department of Chemistry, P.O. Box 55, FI-00014 University of Helsinki, Finland}
\affiliation{University of Jyvaskyla, Department of Physics, P.O. Box 35, FI-40014 Jyvaskyla, Finland}
\author{R.~Julin}
\affiliation{University of Jyvaskyla, Department of Physics, P.O. Box 35, FI-40014 Jyvaskyla, Finland}
\author{S.~Juutinen}
\affiliation{University of Jyvaskyla, Department of Physics, P.O. Box 35, FI-40014 Jyvaskyla, Finland}
\author{J.~Konki}
\altaffiliation{Present address: CERN, CH-1211 Geneva 23, Switzerland}
\affiliation{University of Jyvaskyla, Department of Physics, P.O. Box 35, FI-40014 Jyvaskyla, Finland}
\author{M.~Leino}
\affiliation{University of Jyvaskyla, Department of Physics, P.O. Box 35, FI-40014 Jyvaskyla, Finland}
\author{A.~Lightfoot}
\affiliation{University of Jyvaskyla, Department of Physics, P.O. Box 35, FI-40014 Jyvaskyla, Finland}
\author{A.~Lopez-Martens}
\affiliation{CSNSM, IN2P3-CNRS, F-91405 Orsay Campus, France}
\author{A.~Mistry}
\altaffiliation{Present address: GSI Helmholtzzentrum f\"ur Schwerionenforschung GmbH, 64291 Darmstadt, Germany}
\affiliation{University of Liverpool, Department of Physics, Oliver Lodge Laboratory, Liverpool L69 7ZE, United Kingdom}
\author{P.~Nieminen}
\altaffiliation{Present address: Fortum Oyj, Power Division, P.O. Box 100, 00048 Fortum, Finland}
\affiliation{University of Jyvaskyla, Department of Physics, P.O. Box 35, FI-40014 Jyvaskyla, Finland}
\author{J.~Pakarinen}
\affiliation{University of Jyvaskyla, Department of Physics, P.O. Box 35, FI-40014 Jyvaskyla, Finland}
\author{P.~Papadakis}
\affiliation{University of Jyvaskyla, Department of Physics, P.O. Box 35, FI-40014 Jyvaskyla, Finland}
\affiliation{University of Liverpool, Department of Physics, Oliver Lodge Laboratory, Liverpool L69 7ZE, United Kingdom}
\author{J.~Partanen}
\affiliation{University of Jyvaskyla, Department of Physics, P.O. Box 35, FI-40014 Jyvaskyla, Finland}
\author{P.~Peura}
\affiliation{University of Jyvaskyla, Department of Physics, P.O. Box 35, FI-40014 Jyvaskyla, Finland}
\author{P.~Rahkila}
\affiliation{University of Jyvaskyla, Department of Physics, P.O. Box 35, FI-40014 Jyvaskyla, Finland}
\author{J.~Rubert}
\affiliation{Institut Pluridisciplinaire Hubert Curien, F-67037 Strasbourg, France}
\author{P.~Ruotsalainen}
\affiliation{University of Jyvaskyla, Department of Physics, P.O. Box 35, FI-40014 Jyvaskyla, Finland}
\author{M.~Sandzelius}
\affiliation{University of Jyvaskyla, Department of Physics, P.O. Box 35, FI-40014 Jyvaskyla, Finland}
\author{J.~Saren}
\affiliation{University of Jyvaskyla, Department of Physics, P.O. Box 35, FI-40014 Jyvaskyla, Finland}
\author{C.~Scholey}
\affiliation{University of Jyvaskyla, Department of Physics, P.O. Box 35, FI-40014 Jyvaskyla, Finland}
\author{J.~Sorri}
\altaffiliation{Present address: Sodankylä Geophysical Observatory, University of Oulu, 90014 Oulu, Finland}
\affiliation{University of Jyvaskyla, Department of Physics, P.O. Box 35, FI-40014 Jyvaskyla, Finland}
\author{S.~Stolze}
\altaffiliation{Present address: Physics Division, Argonne National Laboratory, 9700 South Cass Avenue, Lemont, Illinois 60439, USA}
\affiliation{University of Jyvaskyla, Department of Physics, P.O. Box 35, FI-40014 Jyvaskyla, Finland}
\author{B.~Sulignano}
\affiliation{Irfu, CEA, Universit\'e Paris-Saclay, F-91191 Gif-sur-Yvette, France}
\author{J.~Uusitalo}
\affiliation{University of Jyvaskyla, Department of Physics, P.O. Box 35, FI-40014 Jyvaskyla, Finland}
\author{A.~Ward}
\affiliation{University of Liverpool, Department of Physics, Oliver Lodge Laboratory, Liverpool L69 7ZE, United Kingdom}
\author{M.~Zieli\'nska}
\affiliation{Irfu, CEA, Universit\'e Paris-Saclay, F-91191 Gif-sur-Yvette, France}
%

\date{\today}

\begin{abstract}

In the study of the odd-$Z$, even-$N$ nuclei $^{243}$Es and $^{249}$Md, performed at the
University of Jyv\"askyl\"a, the fusion-evaporation
reactions $^{197}$Au($^{48}$Ca,2$n$)$^{243}$Es and
$^{203}$Tl($^{48}$Ca,2$n$)$^{249}$Md have been used for the first
time.  Fusion-evaporation
residues were selected and detected using the RITU gas-filled separator
coupled with the focal-plane spectrometer GREAT.  For $^{243}$Es, the  recoil
decay correlation analysis yielded a half-life of $24 \pm 3$\,s, and a
maximum production cross section of $37 \pm 10$\,nb.  In the same way, a half-life of $26
\pm 1$\,s, an $\alpha$ branching ratio of 75 $\pm$ 5\%, and a maximum production cross section of 300 $\pm$
80\,nb were  determined for $^{249}$Md.
The decay properties of $^{245}$Es, the daughter of $^{249}$Md, were also
measured: an $\alpha$ branching ratio of 54 $\pm$ 7\%  and a half-life of 65 $\pm$ 6\,s.
Experimental cross sections were compared to the results of calculations performed using
the \textsc{kewpie2} statistical fusion-evaporation code.
\end{abstract}

\maketitle	
\section{Introduction}
\label{introduction}
Determining the boundaries of the nuclear chart, particularly in the region of super-heavy nuclei
(SHN), is one of the key questions driving fundamental nuclear physics.
The SHN owe their existence to shell effects as without them the Coulomb repulsion would
make
the nuclei beyond $Z=104$ unstable against fission~\cite{Ackermann2017}.
In this context, detailed spectroscopy of very heavy nuclei (VHN) and SHN is
of paramount importance to provide information on the nuclear landscape
close to
the high-$A$ limit
of the nuclear chart, as well as on the nature of the predicted
island of stability.  The challenge of these experiments is related to
low production cross sections and, in odd-mass nuclei, to the complexity
of spectra where various collective and single-particle excitations may lie
close in energy.  On the other hand, the studies of odd-mass nuclei are rewarded by
the wealth of information regarding single-particle states, exceeding what can be obtained for even-even nuclei~\cite{Asai2015}.

Regarding the known excited states of single-particle or collective nature,
little data is available for Es ($Z=99$) and Md ($Z=101$)
isotopes~\cite{Asai2015,Theisen2015}.
Before in-beam spectroscopy of these odd-$Z$ nuclei can be attempted,
feasibility studies are a prerequisite, in particular, measurements of production cross sections.
Such measurements also help to improve the description of the fusion-evaporation
reaction mechanism, providing new constraints
for the models.

In this paper, the production cross sections for $^{243}$Es and $^{249}$Md
populated directly in the fusion-evaporation reactions
$^{197}$Au($^{48}$Ca,2$n$)$^{243}$Es and
$^{203}$Tl($^{48}$Ca,2$n$)$^{249}$Md are reported.
The targets and projectiles were chosen as a compromise between the predicted production cross sections and the transmission in the separator.
In particular, very asymmetric reactions using actinide targets 
were not considered, as in such cases: (i) The large angular dispersion due to the low recoil velocity and neutron emission
results in a poor transmission, (ii) the low recoil energy reduces the detection efficiency at the focal plane,
both effects being not fully compensated by enhanced cross sections.

The present study also allowed the half-lives and decay properties of these nuclei to be updated as well
as those of $^{245}$Es, populated by the $\alpha$ decay of $^{249}$Md.
It should be noted that $\alpha$-decay branching ratios and, to a lesser extent,
half-lives are needed to deduce production cross sections.
Finally, the measured production
cross sections for $^{243}$Es and $^{249}$Md are discussed
in the context of the $Z \simeq 100$ region and compared to
the predictions
of the \textsc{kewpie2} statistical fusion-evaporation code~\cite{Lu2016}.

\section{Experimental setup}
\label{experiment}
The experiments were performed at the Accelerator Laboratory of the
University of Jyv\"askyl\"a (JYFL).  The fusion-evaporation residues, including $^{243}$Es
and $^{249}$Md, were separated from the fission fragments, the primary
$^{48}$Ca beam and the beam- and target-like reaction products using the
Recoil Ion Transport Unit (RITU) gas-filled
separator~\cite{Leino1995,Saren2011}, which was operated at a He pressure of
0.4\,-\,0.6\,mbar.  The RITU transmission is estimated to be approximately
30\% for the reactions considered here.
The beam current was measured at regular intervals using a Faraday cup
and monitored using the detectors counting rate, thus, allowing the beam dose to be deduced with an uncertainty of 20\%.

At the focal plane of RITU, the separated fusion-evaporation residues were
first detected in a position-sensitive multi-wire proportional counter
(MWPC) and then implanted in two adjacent double-sided silicon strip
detectors (DSSDs), both detectors being part of the Gamma Recoil Electron
Alpha-Tagging (GREAT) spectrometer~\cite{Page2003}.
The MWPC provided a
time-of-flight (ToF) and energy loss ($\Delta E$) measurement, allowing:
(i) selection of the fusion-evaporation residues using a
ToF-$\Delta E$ identification matrix (ii) correlations with
the DSSD, which enable the recoiling residues (coincidence) to be discriminated from the
decay products (anti-coincidence).  Each DSSD is 300\,$\mu$m thick and
consists of $60\times40$ strips with a 1\,mm strip pitch.
The $Y$ side of the DSSD was calibrated using an external mixed $^{239}$Pu, $^{241}$Am, and $^{244}$Cm $\alpha$ source.
An energy offset is  applied to account for the energy loss of the $\alpha$ particle
in the detector entrance window (in the case of an external source) and for the
daughter nucleus recoil (decay from the detector after implantation) so that the resulting energy corresponds to the literature value for the nuclei studied in the present work.
The $X$ side was amplified with a higher gain to measure low energy conversion electrons, and calibrated using an external $^{133}$Ba source.
Signals from all detectors were processed by a trigger-less acquisition
system known as the Total Data Readout (TDR) \cite{Lazarus2001}.  The recoil
decay correlation analysis was performed using the software package
\textsc{grain}~\cite{Rahkila2008}: after a first selection using the ToF-$\Delta E$
identification matrix, the fusion-evaporation residues (recoils) were identified
using the energy of the $\alpha$ particles registered in the same pixel of the
DSSD subsequent to the implantation of a recoil.
The SAGE array~\cite{Pakarinen2014ChT} surrounded the target for the prompt gamma and conversion-electron detection,
however, data from this detector were not used in the present work.

\DeclareRobustCommand{\Es}{$^{243}$Es}
\section{{\Es} decay properties and production cross section}
\label{resEs}
\subsection{Decay and half-life measurement}
\label{resEsDecay}
The $^{243}$Es isotope was discovered  in the 1970s by Eskola {\it et al.} using
the $^{233}$U($^{15}$N,5$n$)$^{243}$Es reaction~\cite{Eskola1972,Eskola1973a}, and later revisited in
the 1990s by Hatsukawa {\it et al.} using the $^{233}$U($^{14}$N,4$n$)$^{243}$Es reaction~\cite{Hatsukawa1989}.
A more recent study, performed with the SHIP separator at GSI by Antalic
{\it et al.}~\cite{Antalic2010}, has shown that  $^{243}$Es decays to its daughter via an
$\alpha$-particle with an energy of
7893 $\pm$ 10\,keV, with a half-life of $T_{1/2}=23 \pm 3$\,s and an $\alpha$-decay branching ratio of $61 \pm 6$\% .
An $\alpha$-particle fine structure was tentatively observed with peaks at 7745 $\pm$ 20 and 7850 $\pm$ 20\,keV.
In the work of Antalic {\it et al.} $^{243}$Es was populated in the decay of the mother
 nucleus $^{247}$Md, whereas in the present study it was directly produced in the $^{197}$Au($^{48}$Ca,2$n$)$^{243}$Es reaction,
with a 21\,pnA $^{48}$Ca beam at $\sim 210$\,MeV energy impinging on a $^{197}$Au target.
The $^{48}$Ca~+~$^{197}$Au reaction has already been studied in the 1990s
by G\"aggeler {\it et al.}~\cite{Gaggeler1989}, however, few spectroscopic data
were available at that time, preventing the discrimination of fusion-evaporation
residues from $2n$ and $3n$ channels.

\begin{figure}[htb]
\begin{center}
\resizebox{0.5\textwidth}{!}{
  \includegraphics{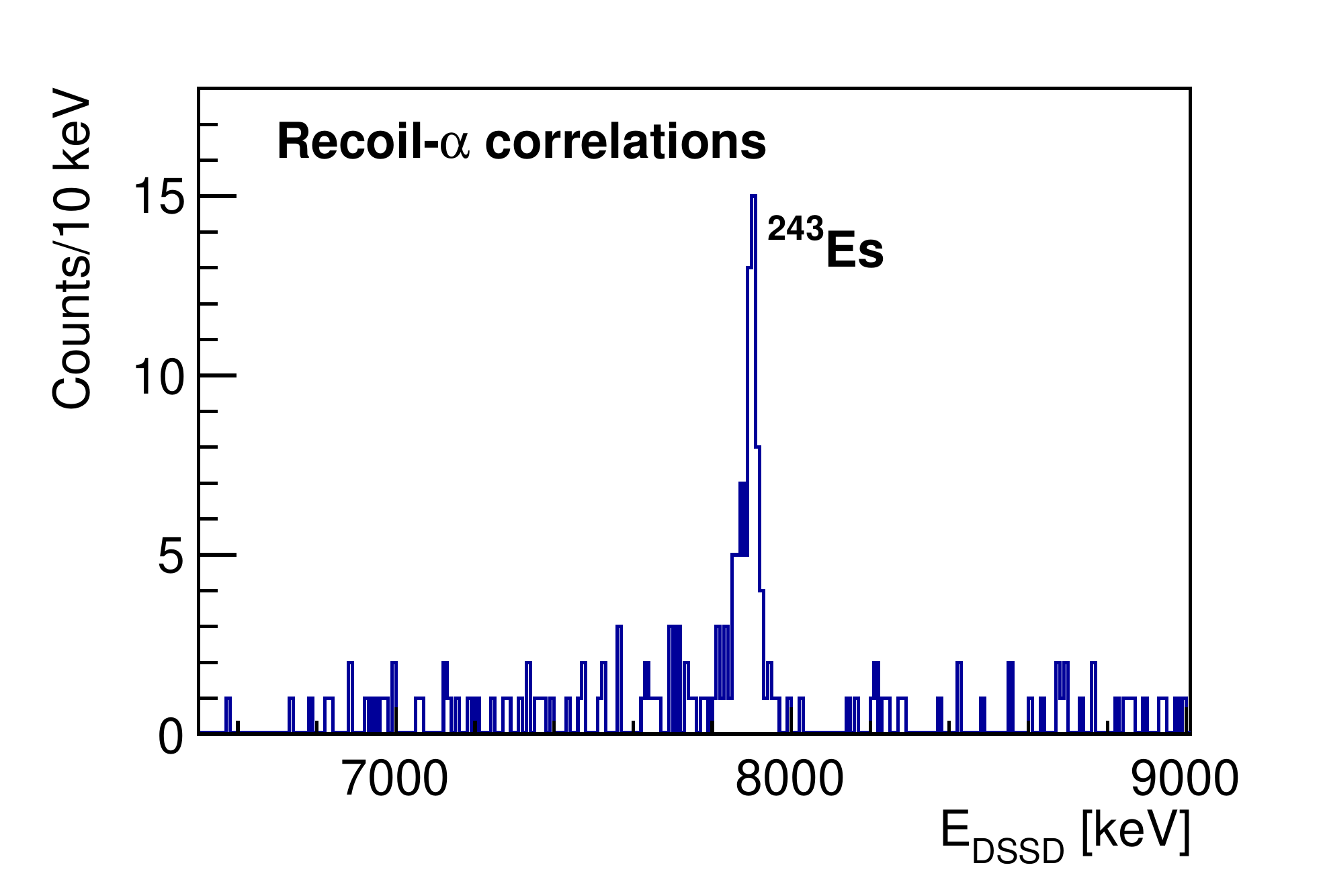}
}
\vspace{-1cm}
\end{center}
\caption{Alpha-particle energy spectrum of $^{243}$Es measured in the DSSD resulting from recoil-$\alpha$ correlations using a maximum search time of 268\,s.}
\label{fig:Es243NAlpha}
\end{figure}

Fig.~\ref{fig:Es243NAlpha} presents the $\alpha$-particle energy spectrum measured in the DSSD resulting from recoil-$\alpha$ correlations,
with the decay of $^{243}$Es
clearly visible.

\begin{figure}[htb]
\begin{center}
\resizebox{0.50\textwidth}{!}{
  \includegraphics{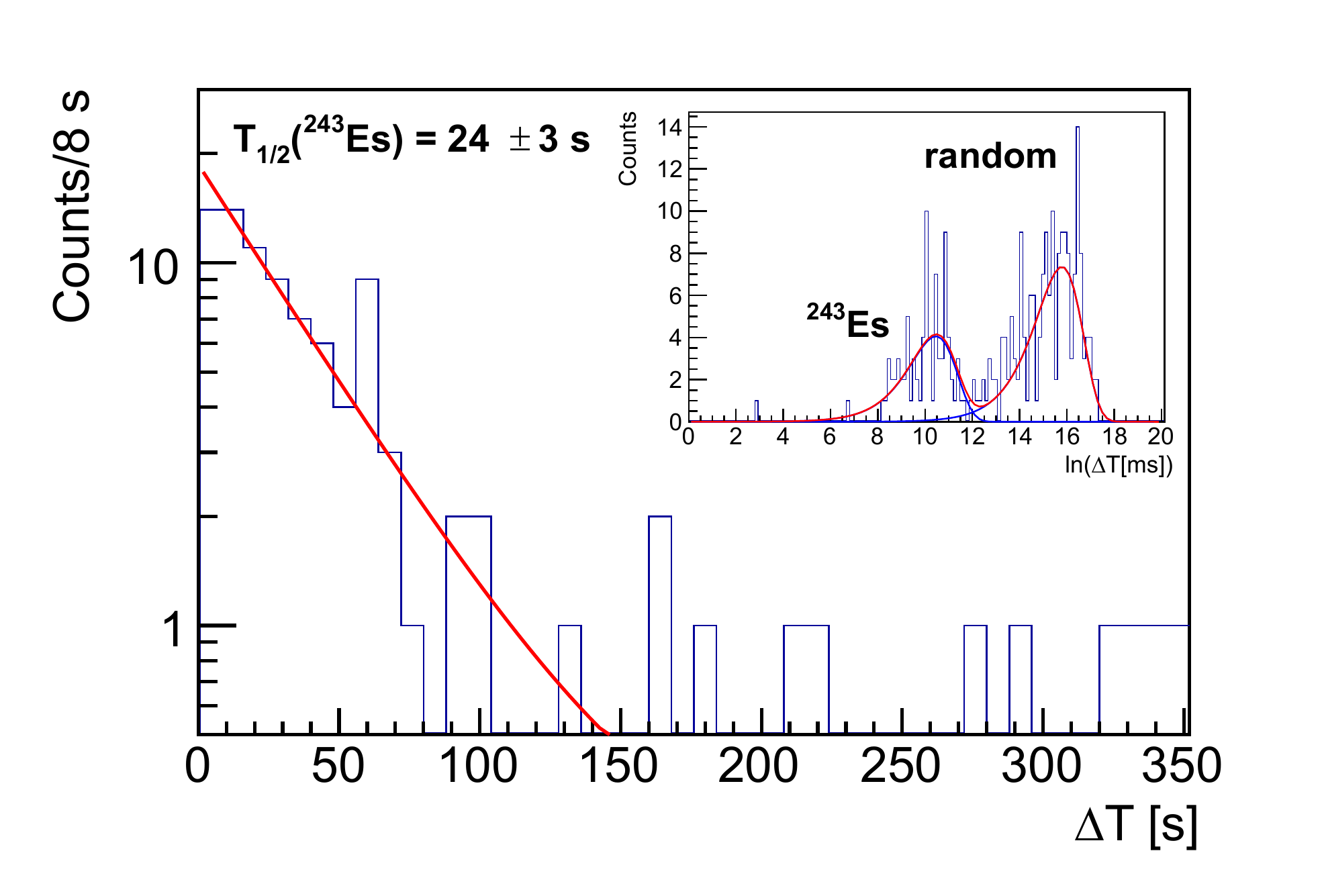}
}
\vspace{-1cm}
\end{center}
\caption{Time distribution of $\alpha$ decays with respect to the $^{243}$Es fusion-evaporation residue implantation. The inset
shows the same data as a function of  $\textrm{ln}(\Delta T)$, with $\Delta T$ expressed in milliseconds.
It should be noted that the range is different for the two spectra: 350 s
for the main panel and 135 h for the inset.
The fit using a two-component decay curve (real and random) is shown with a solid line.}
\label{fig:Es243Halflife}
\end{figure}

The time distribution ($\Delta T$) of the $\alpha$ decay with respect to the implantation, selecting the $^{243}$Es $\alpha$-decay energy, is presented in
Fig.~\ref{fig:Es243Halflife}.
In the inset, the time distribution is drawn 
as a function of $\textrm{ln}(\Delta T)$ using a maximum search time of 10\,h.
The peak at $\textrm{ln}(\Delta T)=10.5$ corresponds to the $^{243}$Es decay,
whereas that around $\textrm{ln}(\Delta T)=16$ is related to random correlations
occurring at an average time interval of $\approx$ 5000\,s.  The spectrum in
the main panel can be fitted using the function~\cite{Leino1981},
\begin{equation}
f(T) = A e^{-(\lambda + r)\Delta T} + B e^{-r \Delta T},
\label{equ.time}
\end{equation}
where $\lambda$ is the decay constant of the nucleus of interest and $r$ is the
random correlation rate.  Similarly, the spectrum in the inset can be fitted
following the method described in Ref.~\cite{Schmidt2000}.  As
expected, both procedures give the same result, yielding the
half-life of $T_{1/2} = 24 \pm 3$\,s, in agreement with the results of the experiment
performed at SHIP~\cite{Antalic2010}.

The inset of Fig.~\ref{fig:Es243Halflife} demonstrates that the
$^{243}$Es decay events can be well separated from the background in the
defined range $\textrm{ln}(\Delta T)<12.5$, which corresponds to a time window of 268\,s
after the recoil implantation.  This search time is used in the
next section in order to determine the number of events corresponding to the $\alpha$ decay
of $^{243}$Es.

The recoil-$\alpha$-$\alpha$ correlations were used to search for the decay of $^{239}$Bk
following the $^{243}$Es decay. The negative outcome of this search is again consistent with
the results of the measurement at SHIP~\cite{Antalic2010}.  The decay properties of nuclei
studied in the present work are summarized in Table~\ref{tabSummary}.


\begin{table}[htb]
\caption{Summary of decay properties obtained in the present work compared to the literature
values.}
\label{tabSummary}
\begin{ruledtabular}
\begin{tabular}{lccc}
Nucleus & Half-life (s) & $\alpha$-decay branching ratio (\%) & Reference \\
\noalign{\smallskip}\hline\noalign{\smallskip}
  $^{243}$Es & 24 $\pm$ 3 & $^{239}$Bk not observed & This work \\
             & 23 $\pm$ 3 & 61 $\pm$ 6 & \cite{Antalic2010} \\
\noalign{\smallskip}\hline\noalign{\smallskip}
  $^{245}$Es & 65 $\pm$ 6 & 54 $\pm$ 7 & This work \\
             & & 40 $\pm$ 10 & \cite{Eskola1973} \\
             & $80^{+96}_{-28}$ & $80^{+20}_{-50}$ & \cite{Hesberger1985} \\
             & 66 $\pm$ 6 & & \cite{Hatsukawa1989} \\
             & $55^{+12}_{-8.4}$ & & \cite{Gates2008} \\
\noalign{\smallskip}\hline\noalign{\smallskip}
  $^{249}$Md & 26 $\pm$ 1 & 75 $\pm$ 5 & This work \\
             & $25^{+14}_{-7}$ & $>$ 60 & \cite{Hesberger1985}  \\
             & $19^{+3}_ {-2}$ & & \cite{Hessberger2001}  \\
             & $23.8^{+3.8}_ {-2.9}$ & & \cite{Gates2008}  \\
             & $23 \pm 3$ & & \cite{Hessberger2009}  \\
             &            & 75\ & \cite{Streicher2006} \\
\noalign{\smallskip}
\end{tabular}
\end{ruledtabular}
\end{table}

\subsection{Production cross section}
\label{resEsSigma}
In order to study the production cross section for $^{243}$Es using the
fusion-evaporation reaction $^{197}$Au($^{48}$Ca,2$n$)$^{243}$Es, two
different beam energies were used.
The target used for this measurement was a $270 \pm 13$\,$\mu$g\,cm$^{-2}$ thick
$^{197}$Au self-supporting foil.
The cyclotron delivered a $213.0 \pm 1.0$\,MeV beam first passing through the 100\,$\mu$g\,cm$^{-2}$ carbon window of the SAGE electron spectrometer.
The first part of the study was performed with a beam energy in the Middle of the Target (MoT) estimated to be $210.0 \pm 1.0$\,MeV.
Then a carbon degrader foil of $100$\,$\mu$g\,cm$^{-2}$ was placed upstream to reduce the incident energy (MoT)
to  $208.0 \pm 1.0$\,MeV.
The spectrum presented in Fig.~\ref{fig:Es243NAlpha} corresponds to the
total statistics, namely with and without
the degrader.

The number of counts attributed to the $^{243}$Es $\alpha$ decay was obtained using a maximum search time of 268\,s.
The contribution from random correlations was estimated by integrating the random correlations component
(second term in Eq.~\ref{equ.time} in the case $\lambda \gg r$) using this time window.
After subtracting this background, the number of $\alpha$ particles stemming from $^{243}$Es
was determined to be 50 $\pm$ 7 (32 $\pm$ 6)
without (with) the carbon degrader foil.
The uncertainties were evaluated following the method described in Ref.~\cite{Bruchle2003}.
In the present work, the statistics is large enough to consider standard normal
distributions, therefore, symmetric uncertainties are
adopted.

During the acquisition time without and with the degrader,
the number of $^{48}$Ca
nuclei that impinged on the $^{197}$Au target was equal to
($1.6 \pm 0.3)
\times 10^{16}$ and ($1.2 \pm 0.2) \times 10^{16}$, respectively.
Taking into account the $^{197}$Au
target thickness, 
the $\alpha$-decay branching ratio of
$61 \pm 6$\% \cite{Antalic2010}, the $\alpha$-detection efficiency of
55\%, and assuming a RITU transmission of 30\%,
a production cross section $\sigma(^{243}$Es$) = 37 \pm 10$\,nb was deduced for a beam energy of $210.0 \pm 1.0$\,MeV  (without the degrader),
and  $\sigma(^{243}$Es$) = 32 \pm 9$\,nb for a beam energy of $208.0 \pm 1.0$\,MeV (with the degrader).
Only statistical uncertainties corresponding to the beam dose, number of $\alpha$-particles, and $\alpha$-decay branching ratio are given.
The RITU transmission of 30\% is actually a transmission $\times$ detection efficiency including the transmission through the separator, the time-of-flight, and the DSSD detection efficiencies.
The results are presented in Table~\ref{tabEs}.

\begin{table}[htb]
\caption{Production cross sections for $^{243}$Es using the
fusion-evaporation reaction $^{197}$Au($^{48}$Ca,2$n$)$^{243}$Es measured
for two different $^{48}$Ca beam energies ($E_{beam}$ corresponds to the middle of target).
$N_\alpha$ is the number of observed $\alpha$ decays after background subtraction.}
\label{tabEs}
\begin{ruledtabular}
\begin{tabular}{cccc}
$E_{beam}$ (MeV) &  $^{48}$Ca dose & $N_\alpha$ & $\sigma$ (nb) \\
\noalign{\smallskip}\hline\noalign{\smallskip}
210.0 $\pm$ 1.0&  $(1.6 \pm 0.3) \times 10^{16}$  & 50 $\pm$ 7 & 37 $\pm$ 10 \\
208.0 $\pm$ 1.0 & $(1.2 \pm 0.2) \times 10^{16}$ & 32 $\pm$ 6 & 32 $\pm$ 9 \\
\end{tabular}
\end{ruledtabular}
\end{table}

\DeclareRobustCommand{\Md}{$^{249}$Md}
\section{{\Md} decay properties and production cross-section}
\label{resMd}
The odd-$Z$ nucleus $^{249}$Md was populated using the fusion-evaporation reaction $^{203}$Tl($^{48}$Ca,2$n$)$^{249}$Md
in three different irradiation campaigns.
The first campaign was focused on cross-section measurements at two different bombarding energies of 214.3 $\pm$ 1.1 and 212.7 $\pm$ 1.1\,MeV.
The results are reported in section~\ref{resMdSigma}.
The two subsequent campaigns aimed principally at the in-beam and decay spectroscopy of $^{249}$Md,
results of which will be reported in a forthcoming paper.
The data collected in the three campaigns were used to derive the $^{245}$Es
and $^{249}$Md half-lives and $\alpha$-decay branching ratios as presented in
the following section.

\subsection{$^{249}$Md and $^{245}$Es decay and half-life measurement}
\label{resMdDecay}
\begin{figure}[htb]
\begin{center}
\resizebox{0.48\textwidth}{!}{
  \includegraphics{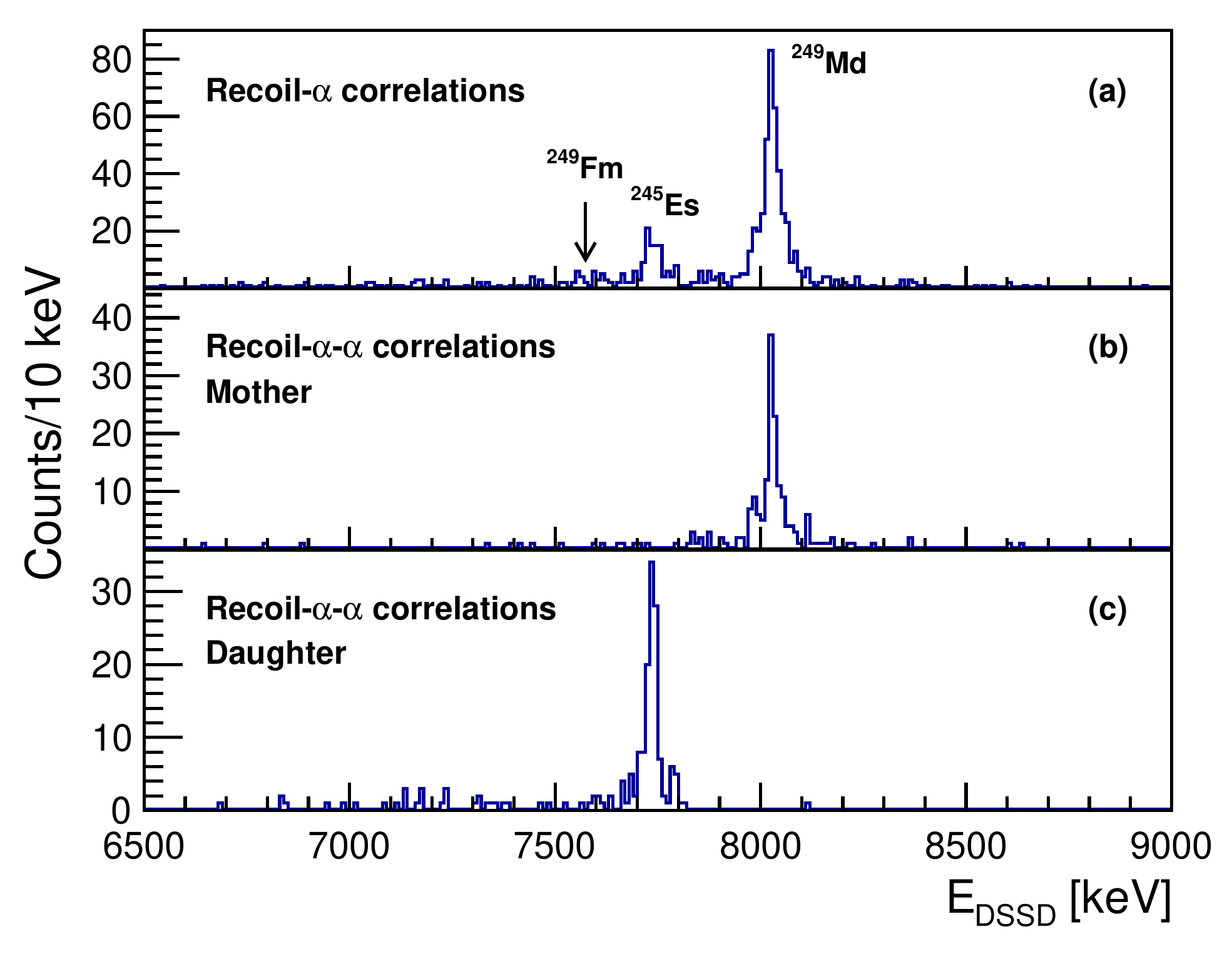}
}
\vspace{-1cm}
\end{center}
\caption{Alpha-particle energy spectra of $^{249}$Md, $^{249}$Fm and $^{245}$Es resulting from recoil-$\alpha$ (a) and
recoil-$\alpha$-$\alpha$ (b),(c) correlations using a maximum search time of
10\,min.}
\label{fig:Md249NAlpha}
\end{figure}

\begin{figure}[htb]
\begin{center}
\resizebox{0.5\textwidth}{!}{
  \includegraphics{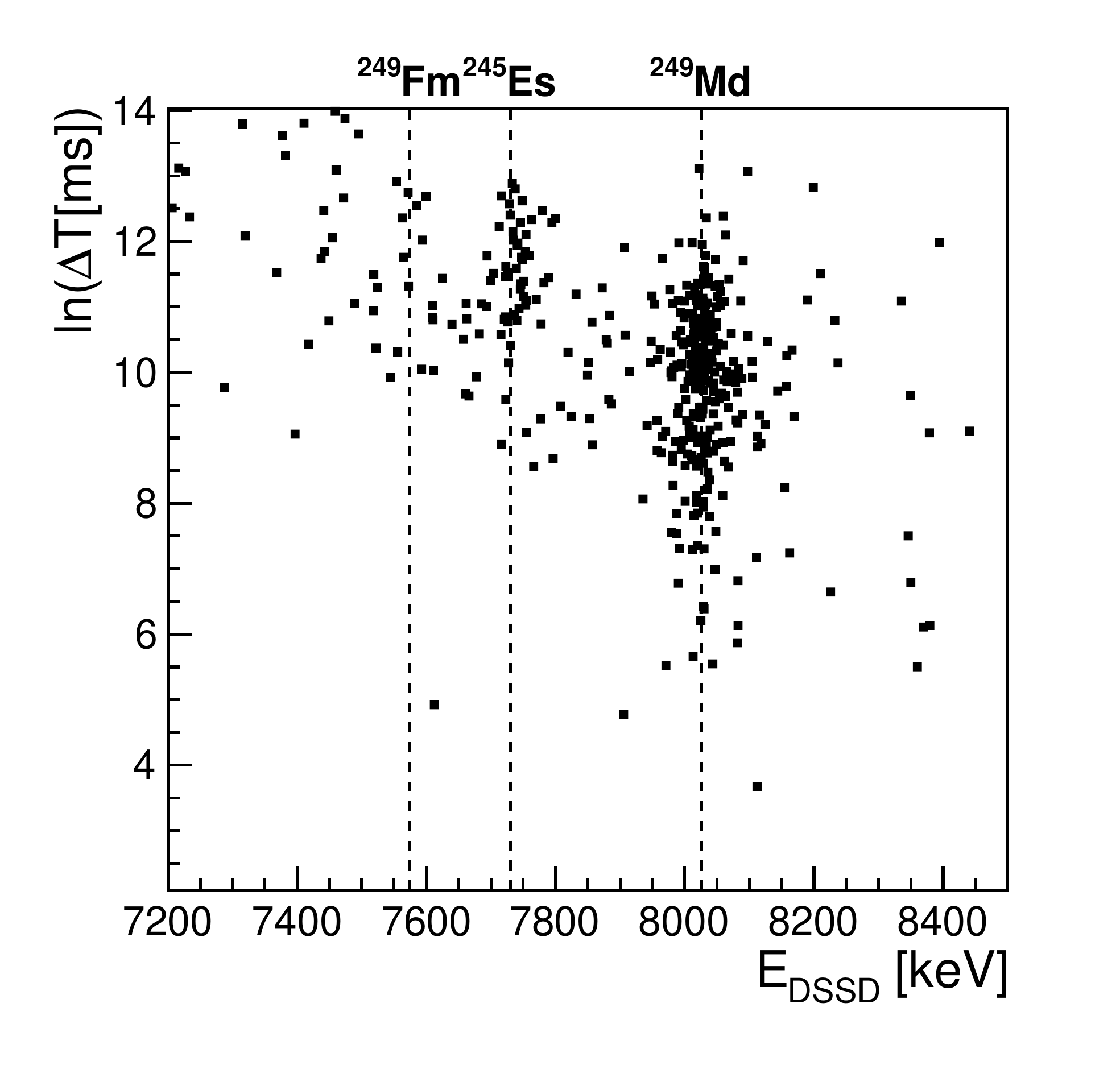}
}
\vspace{-1.5cm}
\end{center}
\caption{Alpha-decay time distribution on a logarithmic scale (from $\sim$ 7\,ms to $\sim$ 20\,min) as a function of the decay energy.}
\label{fig:KHSMd}
\end{figure}

The $\alpha$-particle energy spectra obtained using recoil-$\alpha$ and recoil-$\alpha$-$\alpha$
correlations, with the statistics of the three campaigns
summed together, are presented in Fig.~\ref{fig:Md249NAlpha}.
A maximum search time of 10\,min after the identification of an implanted recoiling nucleus
was used.
$^{249}$Md features an electron capture (EC)/$\beta^+$ decay branch
feeding $^{249}$Fm.  The $\alpha$ decay of the latter is observed using
recoil-$\alpha$ correlations since the detection system is insensitive to the $\beta^+$
particle [see panel (a) of Fig.~\ref{fig:Md249NAlpha}].
The $^{245}$Es $\alpha$ decay observed using recoil-$\alpha$ correlations corresponds to
the events when the $\alpha$ particle emitted from $^{249}$Md escapes from the DSSD
without being detected.
The
$\alpha$ decay of $^{249}$Fm is more clearly visible in
Fig.~\ref{fig:KHSMd},
which represents the $\alpha$-decay time on a logarithmic scale as a
function of the $\alpha$-particle energy.  Using recoil-$\alpha$-$\alpha$
correlations allows the mother, $^{249}$Md and daughter, $^{245}$Es
$\alpha$ decays to be isolated as shown in the (b) and (c) panels of
Fig.~\ref{fig:Md249NAlpha}.
From the literature, the $\alpha$-particle energies are: E$_{\alpha}$($^{249}$Md)$= 8026 \pm
10$\,keV~\cite{Hessberger2009}, and E$_{\alpha}$($^{245}$Es)$= 7730 \pm
1$\,keV~\cite{Hatsukawa1989}.
The satellite peaks in the $\alpha$ decay of $^{249}$Md at 7956 and 8087\,keV, suggested in~\cite{Hessberger2009}, are also tentatively
observed in the present work.

\begin{figure}[htb]
\begin{center}
\resizebox{0.5\textwidth}{!}{
  \includegraphics{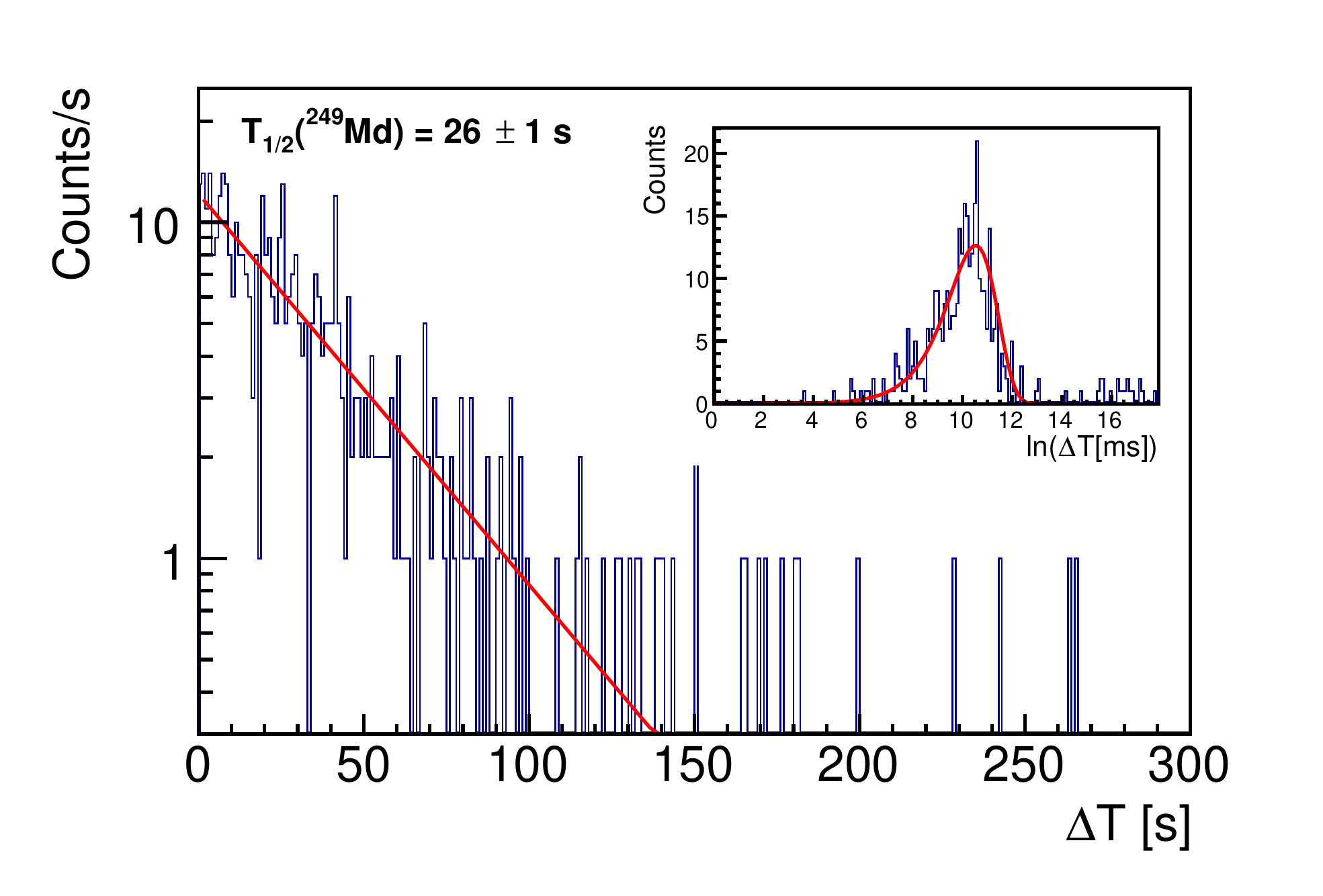}
}
\vspace{-1cm}
\end{center}
\caption{Time distribution of $\alpha$ decays with respect to the $^{249}$Md fusion-evaporation residue implantation. The inset shows
the same data as a function of $\textrm{ln}(\Delta T)$, with $\Delta T$ expressed in milliseconds.
It should be noted that the range is different for the two spectra: 300\,s
for the main panel and 18.2\,h for the inset.
The fit using a one-component decay curve is shown with a solid line.}
\label{fig:Md249Halflife}
\end{figure}

Figure~\ref{fig:Md249Halflife} shows the time distribution of the $^{249}$Md $\alpha$ decay
with respect to the implantation time.
The distribution plotted as a function of $\textrm{ln}(\Delta T)$ for a maximum search time of 24\,h is shown in the inset.
As shown in this plot, the random correlations are negligible, therefore, the time distribution displayed in the main panel can be fitted
with a single exponential function.
A half-life $T_{1/2}=26 \pm 1$\,s is obtained using a maximum search time of 300\,s.
This value can be compared with previously measured half-lives.
The $^{249}$Md decay has been studied at SHIP by Hessberger {\it et al.} following the
$\alpha$ decay of $^{257}$Db ($\rightarrow$ $^{253}$Lr $\rightarrow$
$^{249}$Md)~\cite{Hesberger1985,Hessberger2001} and the $\alpha$ decay of
$^{253}$Lr~\cite{Hessberger2009}, and by Gates {\it et al.} using the Berkeley
Gas-Filled Separator following the $\alpha$ decay of $^{257}$Db~\cite{Gates2008}.
Our revised half-life of $^{249}$Md obtained via direct production and with
higher statistics is compatible
with the values obtained in these works:
$25^{+14}_{-7}$\,s~\cite{Hesberger1985},
$19^{+3}_ {-2}$\,s~\cite{Hessberger2001},
$23 \pm 3$\,s~\cite{Hessberger2009},
$23.8^{+3.8}_ {-2.9}$\,s~\cite{Gates2008}; see also Table~\ref{tabSummary}.

\begin{figure}[htb]
\begin{center}
\resizebox{0.5\textwidth}{!}{
  \includegraphics{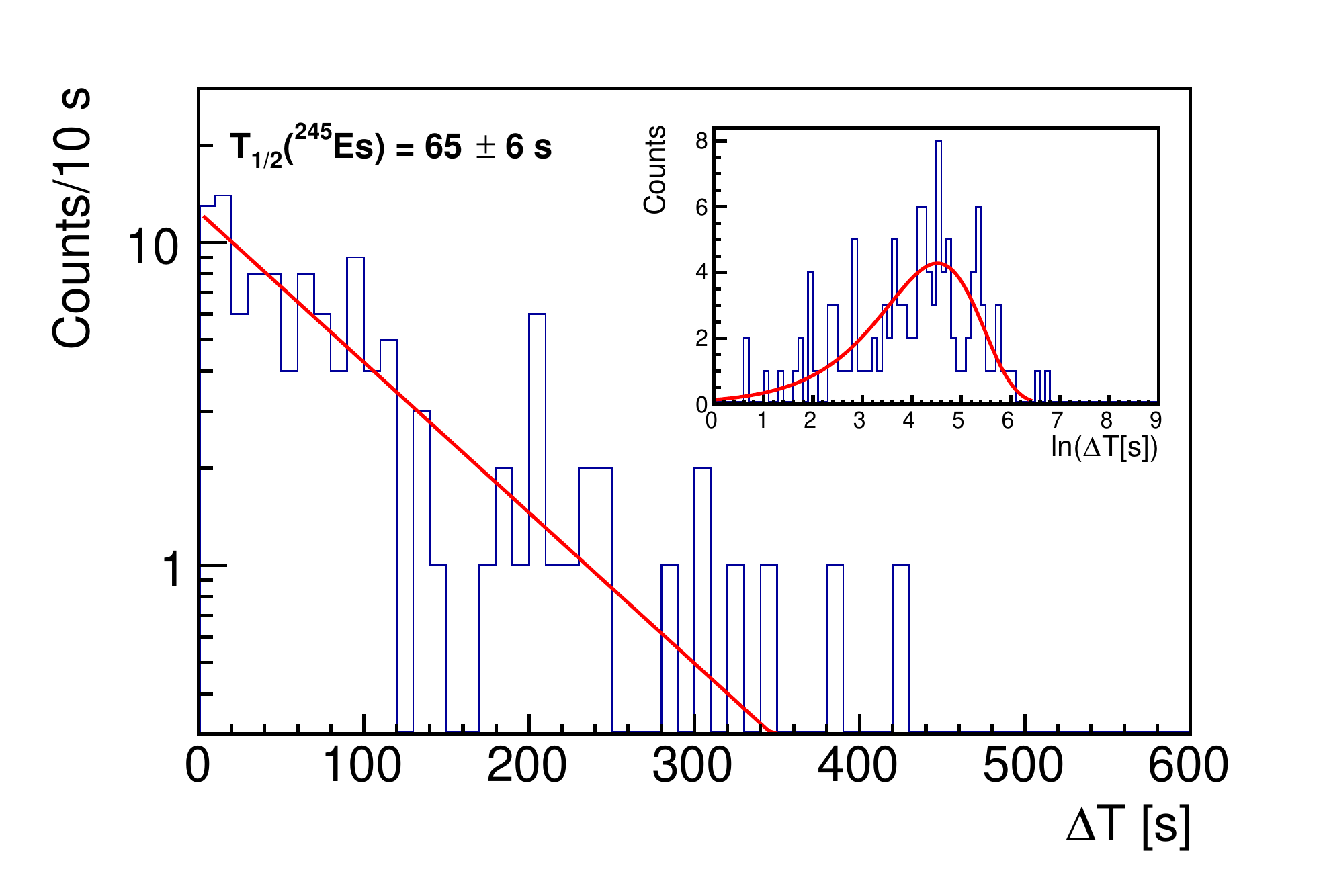}
}
\vspace{-1cm}
\end{center}
\caption{Time distribution of the $\alpha$ decay of $^{245}$Es with respect to the $^{249}$Md $\alpha$ decay. The inset shows
the same data as a function of $\textrm{ln}(\Delta T)$, with $\Delta T$ expressed in seconds.
It should be noted that the range is different for the two spectra: 600\,s
for the main panel and 2.25\,h for the inset.
The fit using a
one-component decay curve is shown with a solid line.}
\label{fig:Es245Halflife}
\end{figure}

Similarly, Fig.~\ref{fig:Es245Halflife} presents the time distribution of
the $^{245}$Es $\alpha$ decay with respect to that of $^{249}$Md, the time
represented in both linear and as a function of $\textrm{ln}(\Delta T)$ scales.
Again, the background is found to be negligible.
The distribution was then fitted with a single component.
The half-life $T_{1/2}(^{245}$Es) = 65 $\pm$ 6\,s was extracted, a value
compatible with those obtained by Hessberger {\it et al.} following the $\alpha$ decay of
$^{257}$Db ($\rightarrow$ $^{253}$Lr $\rightarrow$ $^{249}$Md $\rightarrow$
$^{245}$Es): $80^{+96}_{-28}$ s~\cite{Hesberger1985}, by Hatsukawa {\it et al.}
after direct synthesis using the fusion-evaporation reactions
$^{238}$U($^{14}$N,7$n$)$^{245}$Es and $^{237}$Np($^{12}$C,4n)$^{245}$Es:
66 $\pm$ 6 s~\cite{Hatsukawa1989}, and by Gates {\it et al.} following the $\alpha$ decay
of $^{257}$Db: $55^{+12}_{-8.4}$ s~\cite{Gates2008}; see also Table~\ref{tabSummary}.

The $\alpha$-decay branching ratio of $^{249}$Md is defined as the ratio of the
$\alpha$-decay branch to $^{245}$Es, to the total decay strength, including the
EC/$\beta^+$ branch to $^{249}$Fm.
The latter is evaluated using the number of events attributed to the $^{249}$Fm $\alpha$ decay from
Figs.~\ref{fig:Md249NAlpha}
and~\ref{fig:KHSMd}, corrected for the $^{249}$Fm $\alpha$-decay branching ratio.
A correction is also applied to take into account the fraction of $^{249}$Fm
nuclei that decay during the search time of 600\,s.
The $^{249}$Fm half-life of 2.6 $\pm$ 0.7\,min is taken from the evaluated data~\cite{Abusaleem2011}.
The $^{249}$Fm $\alpha$-decay branching ratio of 15.6 $\pm$ 1.0\% is taken
from Hessberger {\it et al.}~\cite{Hessberger2012},
which is more recent than the evaluation of Ref.~\cite{Abusaleem2011}\footnote{It should be noted that, in Ref.~\cite{Hessberger2012}, the half-life of
$^{249}$Fm has not been re-measured. The value adopted in this reference is actually that of
the evaluation Ref.~\cite{Abusaleem2011}, i.e., 2.6 $\pm$ 0.7\,min. In the most recent \textsc{nubase2016} evaluation~\cite{Audi2017}, the $\alpha$-decay branching ratio of
$^{249}$Fm is taken from Ref.~\cite{Abusaleem2011} (33 $\pm$ 9\%), whereas for the half-life only the value from \cite{Hesberger2004} (96 $\pm$ 6\,s) is selected.
}.
The resulting $\alpha$-decay branching ratio deduced in the present work is $b_{\alpha}(^{249}$Md) = 75 $\pm$ 5\%.
The evaluated value of $b_{\alpha}(^{249}$Md)$>$ 60\%~\cite{Abusaleem2011} corresponds  to
the measurement of Hessberger {\it et al.}, which has been obtained in the study of the $^{257}$Db decay chain~\cite{Hesberger1985}.
A more recent value of $b_{\alpha}(^{249}$Md) = 75\%, quoted without uncertainty
in the PhD thesis of Streicher~\cite{Streicher2006},
is
in perfect agreement with our measurement; see also Table~\ref{tabSummary}.

The $\alpha$-decay branching ratio of $^{245}$Es can be extracted in two distinct ways.
The first possibility is to derive it as the ratio of the number of events
corresponding to $^{249}$Md obtained using
recoil-$\alpha$-$\alpha$ and recoil-$\alpha$ correlations, corrected for
the DSSD efficiency for a full-energy measurement $\epsilon_{\alpha}$ =
55\% under the condition that the recoil-$\alpha$-$\alpha$ correlations are
obtained by gating on the full-energy peaks only,
\begin{equation}
b_{\alpha}(^{245}\textrm{Es}) = \frac{N_{\textrm{recoil}-\alpha-\alpha}(^{249}\textrm{Md})}{N_{\textrm{recoil}-\alpha}(^{249}\textrm{Md})}
\frac{1}{\epsilon_{\alpha}}.
\end{equation}
The second option is to obtain it as the ratio of counts corresponding to  $^{245}$Es and $^{249}$Md
in the total $\alpha$-particle spectrum.
Both methods lead to the same value of $b_{\alpha}(^{245}$Es) = 54 $\pm$ 7\%.
For comparison, the previously reported values were $b_{\alpha}(^{245}$Es) =
40 $\pm$ 10\% (Eskola~\cite{Eskola1973}), $b_{\alpha}(^{245}$Es) =
$80^{+20}_{-50}$\% (Hessberger {\it et al.}~\cite{Hesberger1985}).  The decay
properties of $^{249}$Md and $^{245}$Es are summarized in
Table~\ref{tabSummary}.

\subsection{Production cross section}
\label{resMdSigma}
The fusion-evaporation reaction $^{203}$Tl($^{48}$Ca,2$n$)$^{249}$Md was
studied at two different bombarding energies.
The cyclotron delivered a 218\,MeV beam first passing through the 100 $\mu$g\,cm$^{-2}$ carbon window of the SAGE electron spectrometer.
The $^{203}$Tl target having a thickness of 318 $\pm$ 16\,$\mu$g\,cm$^{-2}$ was evaporated on a carbon foil of 20\,$\mu$g\,cm$^{-2}$, and
covered by a 10\,$\mu$g\,cm$^{-2}$ carbon protection layer.
The resulting energy in the middle of the $^{203}$Tl target was estimated to be 214.3 $\pm$
1.1\,MeV.
Using in addition an 80 $\mu$g\,cm$^{-2}$ carbon degrader foil resulted in an energy
of 212.7 $\pm$ 1.1\,MeV MoT.

The spectra were obtained using a search time of 207\,s, i.e., eight $^{249}$Md half-lives.
Contrary to the $^{243}$Es case, the background was found
to be negligible.

The total number of $^{48}$Ca particles that impinged on the target was  $(1.8 \pm 0.4) \times 10^{15}$ ($(1.5 \pm 0.3) \times 10^{15}$)
for the measurement without (with) the carbon degrader foil.
Using a $^{203}$Tl target thickness of 318 $\pm$ 16\,$\mu$g\,cm$^{-2}$, an
$\alpha$ branching ratio of 75 $\pm$ 5\%, a RITU transmission $\times$ detection efficiency of 30\%
and a full-energy  $\alpha$-detection efficiency of 55\%, cross sections
$\sigma(^{249}$Md) of 300 $\pm$ 80\,nb and 70 $\pm$ 40\,nb are deduced for
the incident energies of 214.3 and 212.7\,MeV, respectively.
Again, only statistical uncertainties are given. The results are
summarized in
Table~\ref{tabMd}.

\begin{table}[htb]
\caption{Production cross sections for  $^{249}$Md using the fusion-evaporation reaction $^{203}$Tl($^{48}$Ca,2$n$)$^{249}$Md measured for two different $^{48}$Ca beam energies ($E_{beam}$
corresponds to the middle of the target). }
\label{tabMd}
\begin{ruledtabular}
\begin{tabular}{cccc}
$E_{beam}$ (MeV) &   $^{48}$Ca dose & $N_\alpha$ & $\sigma$ (nb) \\
\noalign{\smallskip}\hline\noalign{\smallskip}
214.3 $\pm$ 1.1 &  $(1.8 \pm 0.4) \times 10^{15}$  & 63 $\pm$ 8 & 300 $\pm$ 80 \\
212.7 $\pm$ 1.1 & $(1.5 \pm 0.3) \times 10^{15}$  & 12 $\pm$ 4 &  70 $\pm$ 40 \\
\end{tabular}
\end{ruledtabular}
\end{table}

\section{Discussion}

In this section we discuss the new cross-section measurements for $^{243}$Es and $^{249}$Md.
These results are placed in the context of experimental cross sections for
cold fusion-evaporation reactions and the 2$n$ channel for  $Z \approx  100$,
presented in Fig.~\ref{fig:systZ}
and compared to new reactions dynamics calculations using the
statistical fusion-evaporation code \textsc{kewpie2}~\cite{Lu2016}.

\subsection{2$n$ channel fusion-evaporation systematics}
\begin{figure}[hbpt]
\begin{center}
\resizebox{0.53\textwidth}{!}{
  \includegraphics{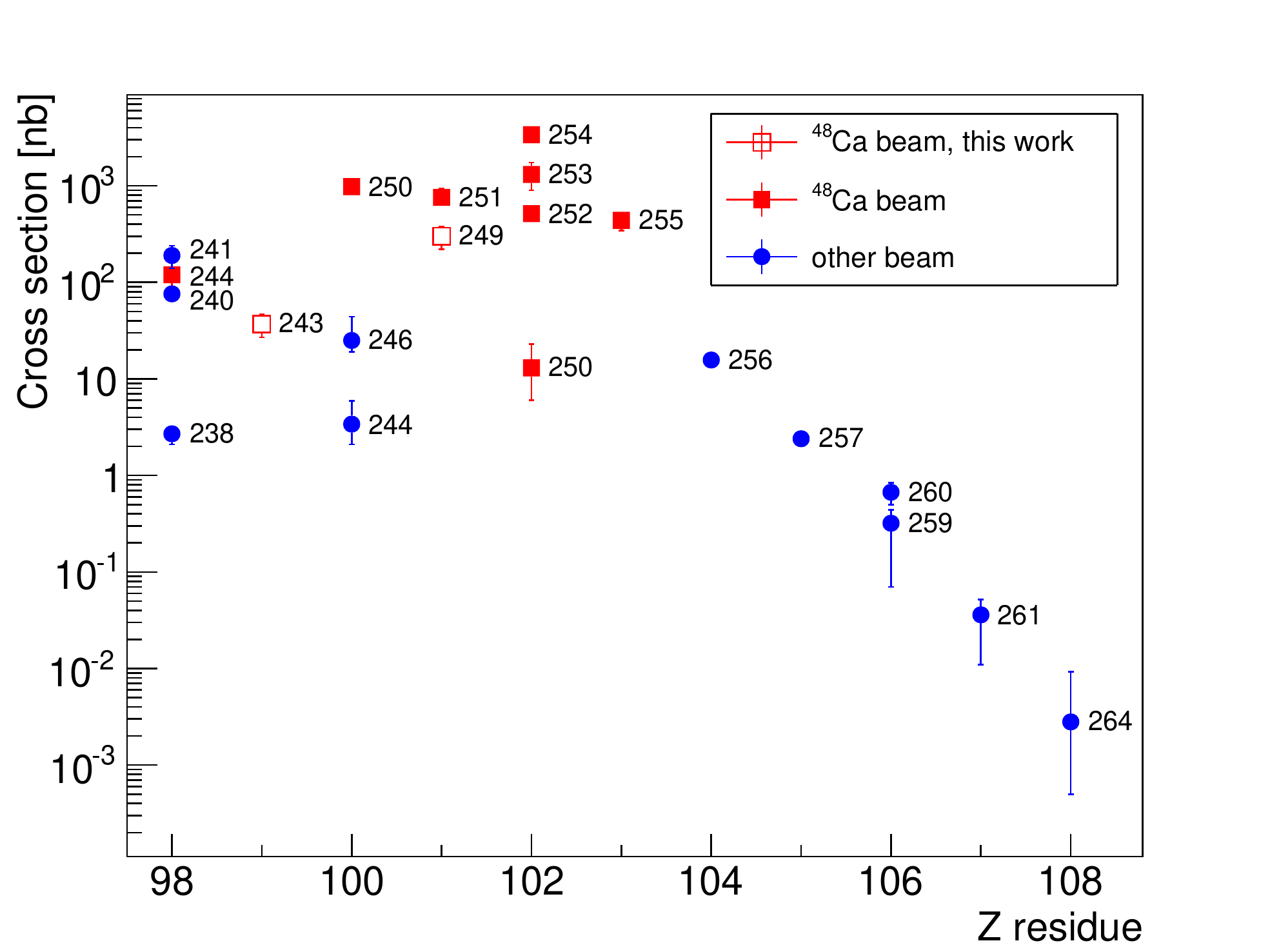}
}
\vspace{-1cm}
\end{center}
\caption{Systematics of the fusion-evaporation cross sections in the 2$n$ channel as a function of $Z$ of the residual nucleus.
The filled red squares correspond to reactions induced by a $^{48}$Ca beam, whereas the
blue circles correspond to those using other beams.
The new $^{243}$Es and $^{249}$Md measurements are denoted by empty square symbols.
The mass number $A$ of the residual nucleus is given to the right of each symbol.
Data are taken from Refs.~\cite{Khuyagbaatar2010} ($^{238,240,241}$Cf),
~\cite{Konki2018} ($^{244}$Cf),
this work ($^{243}$Es, $^{249}$Md),
\cite{Gaeggeler1984} ($^{244,246}$Fm),
\cite{Bastin2006ChT} ($^{250}$Fm),
\cite{Chatillon2007ChT} ($^{251}$Md),
\cite{Peterson2006} ($^{250}$No)
\cite{Oganessian2001} ($^{252,253}$No)
\cite{Gaggeler1989} ($^{254}$No, $^{255}$Lr),
\cite{Drago2008} ($^{256}$Rf),
\cite{Hessberger2001} ($^{257}$Db),
\cite{Muenzenberg1985} ($^{259}$Sg),
\cite{Hessberger2009} ($^{260}$Sg),
\cite{Muenzenberg1989} ($^{261}$Bh),
\cite{Sato2011} ($^{264}$Hs).}
\label{fig:systZ}
\end{figure}

It is generally acknowledged that the fusion-evaporation reactions can be described as three
subsequent independent
processes: capture, compound-nucleus formation, and survival of the residual nucleus.
The description of the capture step is rather well controlled in terms of barrier penetration,
with no rapid evolution as a function of mass and charge when using similar projectiles and targets.
The formation step results in a sharp decrease in the cross section for
projectile-target combinations with $Z_p Z_t \gtrsim 1600-1800$, known as the
fusion hindrance,
which prevents the formation of a compound nucleus by leading the di-nuclear composite towards the quasi-fission route.
This effect starts to
act
in the region considered here, and it can account for
the exponential decrease in the cross sections observed for larger $Z$ values in Fig.~\ref{fig:systZ}.
Consequently, only the survival step can account for the decrease in cross sections below $Z \approx 102$.
The global trend displayed by the cross sections presented in Fig.~\ref{fig:systZ} may be explained by
a combination of two effects.
First, the four-fold magic character of the $^{48}$Ca + $^{208}$Pb
$\rightarrow$ $^{256}$No$^*$ reaction leads to a low $Q$ value and, therefore,
a higher survival probability in the evaporation and de-excitation processes.
This enhancement is observed for $^{254}$No and neighboring residual
nuclei.
Second, the semi-magicity at $Z = 100$, $N = 152$ leads to higher shell corrections
(higher fission barrier) and, therefore, a higher survival probability around $^{252}$Fm.
Note that if the cross sections are plotted as a function of the mass or neutron
number, they also display a bell-shaped behavior.

\subsection{Cross-section calculations}

In the following, the fusion-evaporation cross sections illustrated with the new
experimental results for $^{243}$Es and $^{249}$Md are discussed
in terms of survival from the compound to the residual nucleus, with an emphasis on
the effect of the fission barrier.
The present measurements are performed in a mass region where the fusion hindrance is not yet significant.
Consequently, the fusion process is modelled in the \textsc{kewpie2} code by considering only the capture phase,
which is computed using a proximity potential and
the Wentzel-Kramers-Brillouin (WKB) approximation, see Ref.~\cite{Lu2016} for details.

\begin{figure}[bp]
\vspace{-0.5cm}
\begin{center}
\resizebox{0.5\textwidth}{!}{
 \includegraphics{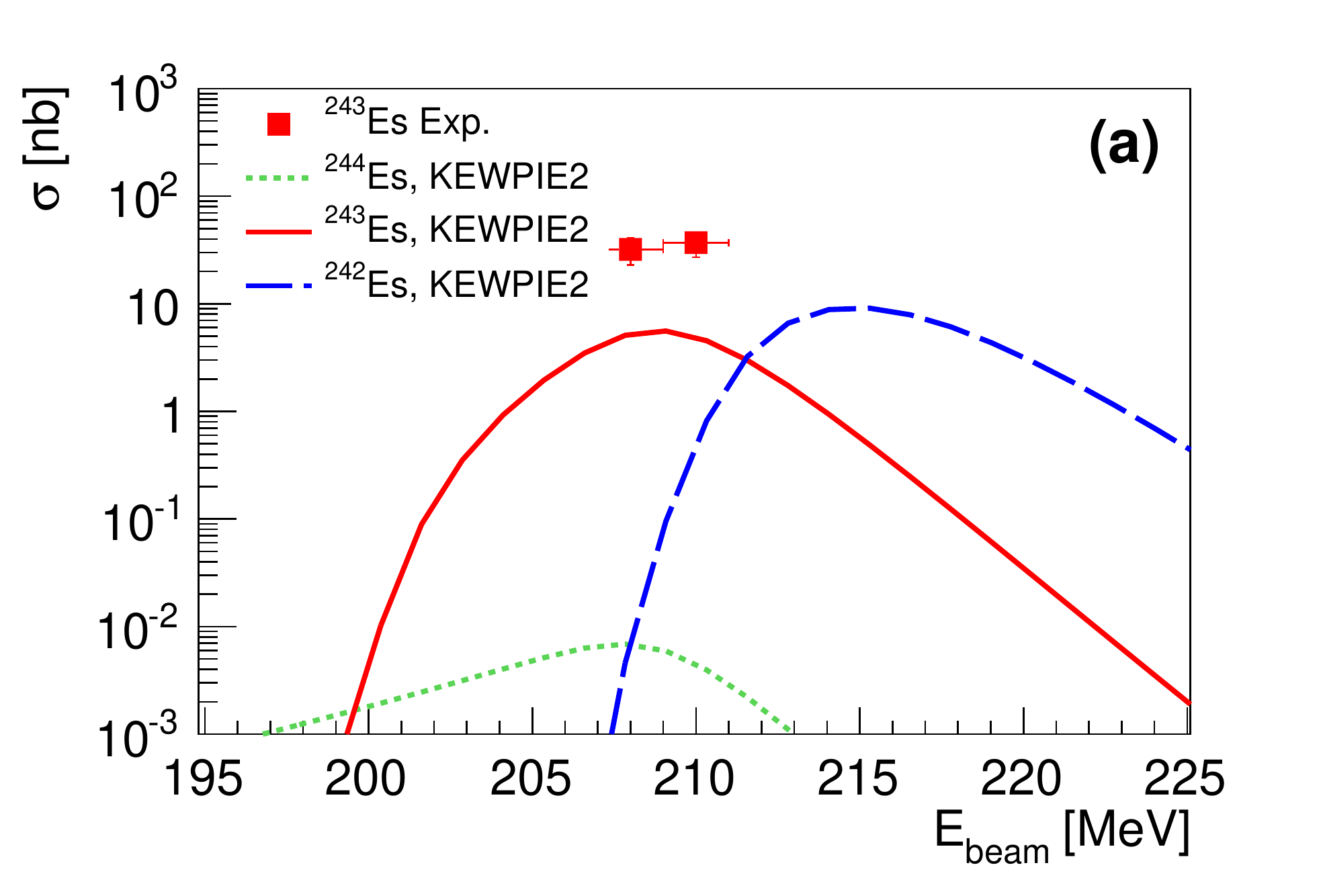}
 }
\resizebox{0.5\textwidth}{!}{
 \includegraphics{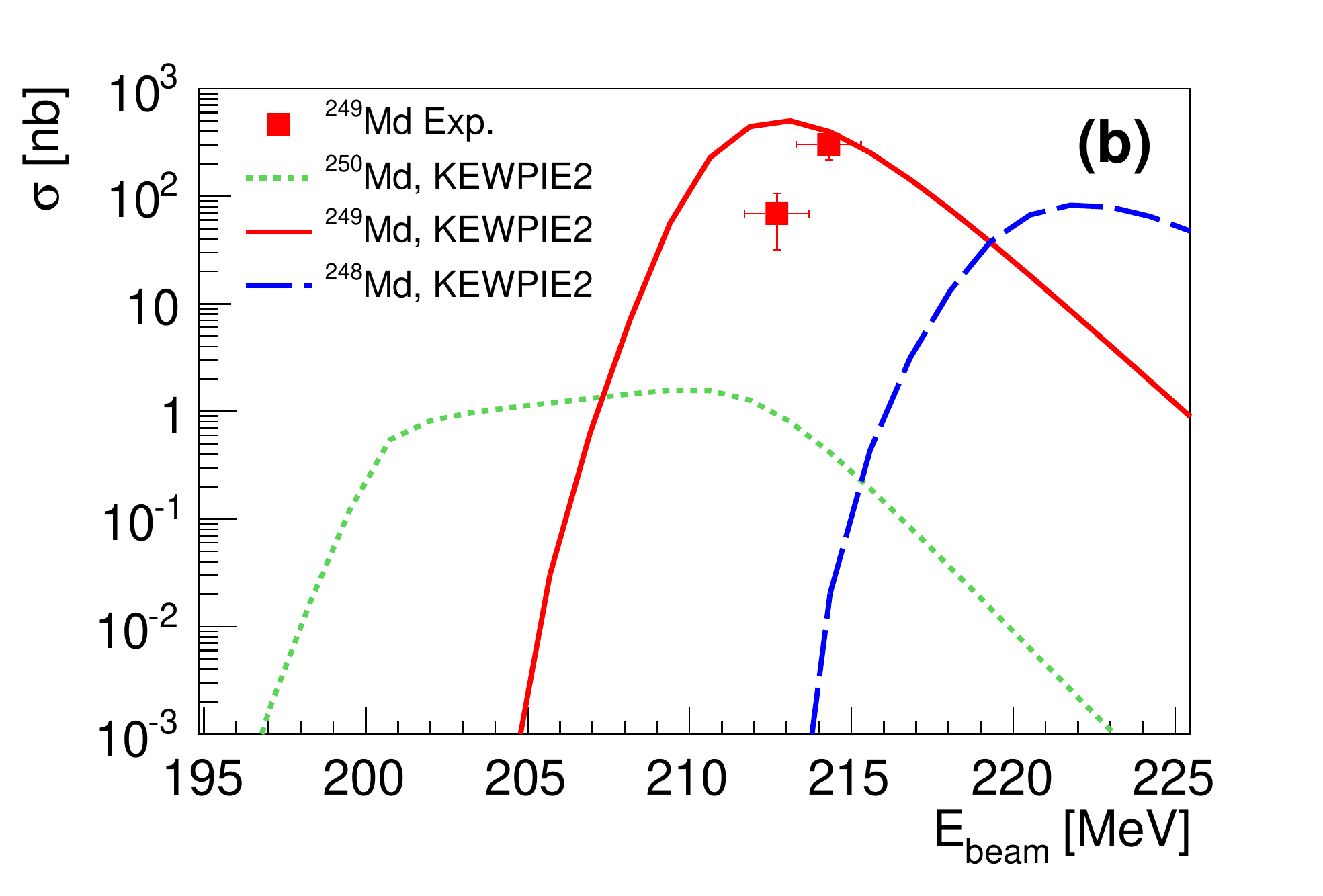}
}
\end{center}
\vspace{-0.5cm}
\caption{Comparison between the experimental production cross sections for $^{243}$Es (a)
obtained in the present work, and
the calculations of the 1$n$, 2$n$, and 3$n$ cross sections performed with the \textsc{kewpie2} code using the default parameters (macroscopic part described by the Thomas-Fermi
parametrization as proposed by Myers-Swiatecki~\cite{Myers1999}
and the microscopic part based on the FRDM shell
corrections~\cite{Moller1995}).
(b) The same for $^{249}$Md.
}
\label{figCompTh}
\end{figure}

The \textsc{kewpie2} code~\cite{Lu2016} treats the competition between light-particle evaporation and fission,
which occurs within an excited compound nucleus using
the statistical formalisms of
Weisskopf~\cite{Weisskopf1937} and Bohr-Wheeler~\cite{Bohr1939},
respectively.
The entire set of default parameters used in the \textsc{kewpie2} code is
presented in Ref.~\cite{Lu2016}.
In the following we will only focus on a few parameters, which are not well-defined either theoretically or experimentally in this mass region~\cite{Lu2016b}.
These parameters are the reduced friction parameter $\beta$, the shell-damping energy $E_d$, and the shell corrections $\Delta E_{\textrm{sh}}$.
These parameters are related to the viscosity of nuclear
matter, the stability of shell corrections with temperature and the fission-barrier
height, respectively, following Eq.~\ref{EqBarrier} for the latter:
\begin{equation}
B_{f} = B_\textrm{LDM} - \Delta E_{\textrm{sh}},
\label{EqBarrier}
\end{equation}
where $B_f$ is the fission-barrier height and $B_\text{LDM}$ is the liquid-drop fission barrier.
The default values used in the \textsc{kewpie2} code are $\beta=2 \times 10^{21}$\,s$^{-1}$,
$E_d=19$\,MeV, whereas the finite-range droplet model (FRDM) $\Delta E_{\textrm{sh}}$ shell
corrections are taken from Ref.~\cite{Moller1995}.
It should  be stressed that those parameters mainly affect the fission
process that is known to be dominant for heavy and super-heavy nuclei.
Indeed, a small variation of the fission parameters, such as the strength
of the dissipation or the fission-barrier heights, leads to a significant
modification of the survival
probability and, consequently, the related observables, in particular the production cross
sections.

\begin{figure}[hbtp]
\begin{center}
\resizebox{0.5\textwidth}{!}{
 \includegraphics{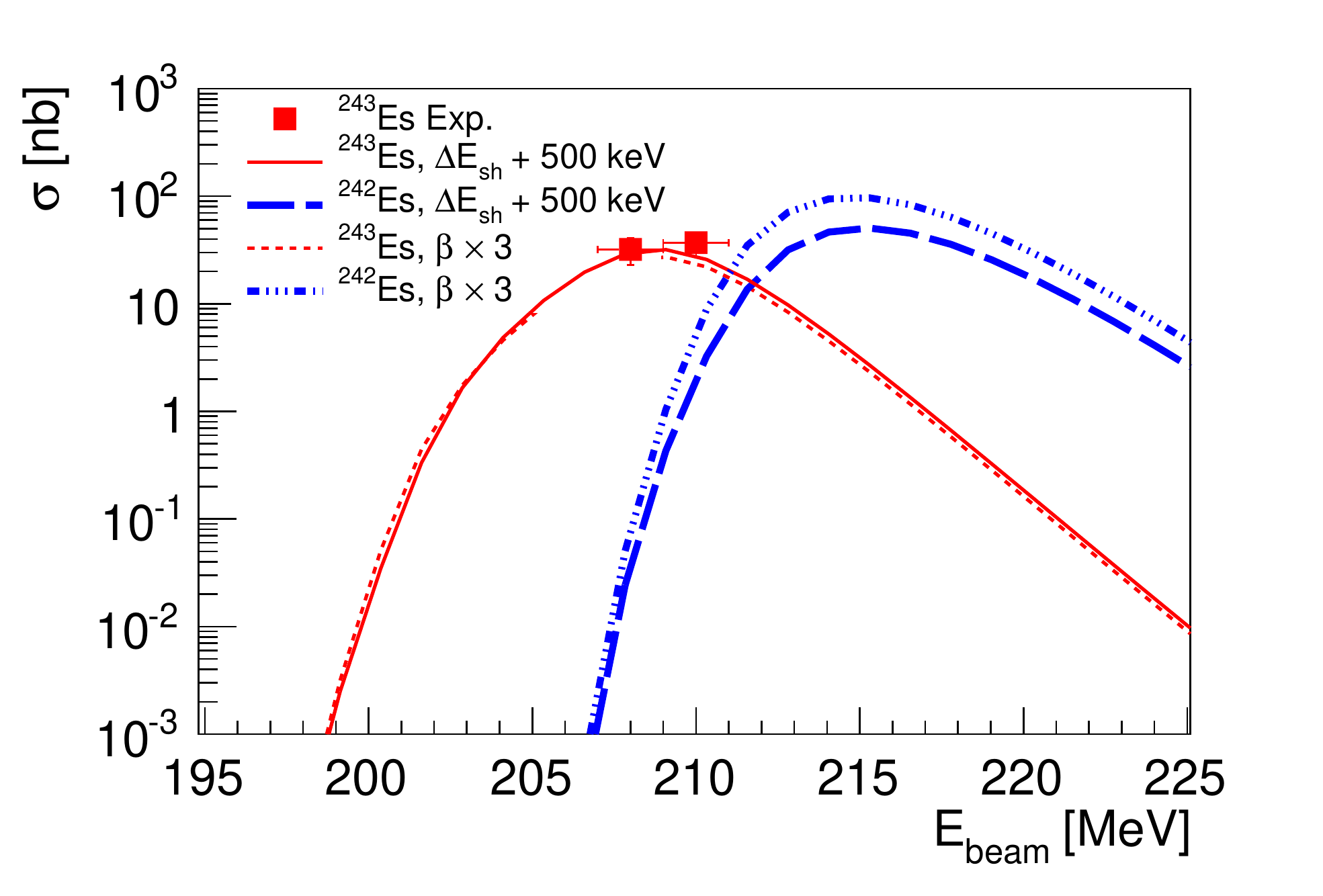}
 }
\end{center}
\vspace{-0.5cm}
\caption{Comparison between the experimental production cross sections for $^{243}$Es
extracted in the present work, and the calculations
performed
with the \textsc{kewpie2} code considering either an adjustment of $+ 500$\,keV of the
barrier heights, or an adjustment of the reduced friction parameter to $\beta
= 6 \times 10^{21}$\,s$^{-1}$.}
\label{figCompThCorr}
\end{figure}

Figure~\ref{figCompTh} presents the experimental
results for the production cross sections for $^{243}$Es and
$^{249}$Md (Tables~\ref{tabEs} and \ref{tabMd}) compared to  the calculations performed with the
\textsc{kewpie2} code using the default parameters.
For $^{249}$Md, the calculation reproduces the measured production
cross sections well, whereas it underestimates them by a factor of 5 for the
$^{243}$Es case.
The discrepancy for this latter case cannot be explained by a failure of the fusion model.
Indeed, for a beam energy corresponding to the present measurement ($E_{cm}\approx169$\,MeV), the fusion model provides a fusion cross section $\sigma_\textrm{fus} = 55$\,mb in good agreement with the measurement $\sigma_\textrm{fus} = 42$\,mb of Ref.~\cite{Pacheco1992}.
Moreover, a discussion of the fusion cross-section for the $^{48}$Ca+$^{208}$Pb reaction, for which the WKB approximation provides a good description without fusion hindrance considerations, can be found in Ref.~\cite{Lu2016}.
In Fig.~\ref{figCompThCorr}, the fission-barrier heights or the reduced
friction parameters have been increased in order to reproduce the
measurements for the $2n$ evaporation channel.
Concerning the fission-barrier heights, it is necessary to add 500\,keV to the
absolute value of the shell corrections [with the liquid-drop fission
barrier kept unchanged, see Eq.~\ref{EqBarrier}], to obtain good agreement
between the calculations and the data.
Furthermore, the reduced friction parameter has to be increased by a
factor of three, i.e.,  to $\beta = 6 \times 10^{21}$\,s$^{-1}$, in order to obtain the same
agreement.
It should be stressed that these adjustments remain within the uncertainty
intervals for these parameters, as discussed in Refs.~\cite{Lu2016, Lu2016b}.
Moreover, no theoretical model can presently predict the fission-barrier heights with
an accuracy better than $0.5-1$\,MeV~\cite{Fritz2017, Capote2009, Kowal2010}.
In the SHN region, differences between the models can be as large as 4\,MeV~\cite{Baran2015ChT}.
Consequently, we cannot attribute the discrepancy observed for $^{243}$Es
(Fig.~\ref{figCompTh}) to any specific parameters used in the
\textsc{kewpie2} code nor to any inputs from other nuclear models, in particular,
those related to  the fission process.
Hence, the measured production cross sections for the $^{243}$Es and
$^{249}$Md isotopes can be fully explained within the uncertainties in nuclear
models and phenomenological parametrizations implemented in the \textsc{kewpie2}
code.

A way to provide constraints on the parameters used in the \textsc{kewpie2} code would be to use more precise measurements in the VHN and SHN mass regions
for a whole set of different evaporation channels, including a large scan in excitation energy for each of them.
Indeed, using relevant data can help to fix and/or eliminate the impact of a specific parameter.
Such an approach based on the Bayesian inference is discussed in~\cite{Lu2016,LueHongliang2013}.

\section{Conclusion}
The odd-$Z$
$^{243}$Es and $^{249}$Md were produced in the $^{197}$Au($^{48}$Ca,2$n$)$^{243}$Es and $^{203}$Tl($^{48}$Ca,2$n$)$^{249}$Md fusion-evaporation reactions, respectively.
The half-life of $^{243}$Es, $^{249}$Md and its daughter $^{245}$Es were measured,
and the results were found compatible with those obtained in previous measurements
following $\alpha$-decay of heavier nuclei.
The precisions of the half-lives of  $^{249}$Md and $^{245}$Es were increased
as well as those of the
$\alpha$-decay branching ratios for those nuclei.

Production cross-sections of $^{243}$Es and $^{249}$Md have been measured
for the first time using $^{48}$Ca-induced reactions and compared to
the calculations performed with the \textsc{kewpie2} code~\cite{Lu2016}.


\begin{acknowledgments}
We acknowledge the accelerator staff at the University of Jyv\"askyl\"a for delivering a high-quality beam during the experiments.
The Au targets were kindly provided by GSI.
Support has been provided by
the
EU 7th Framework Programme “Integrating Activities - Transnational Access” Project No.
262010 (ENSAR) and by the Academy of Finland under the Finnish Centre of Excellence
Programme (Nuclear and Accelerator Based Physics Programme at JYFL contract No. 213503).

\end{acknowledgments}

\bibliographystyle{myapsrev4-1new}
\bibliography{paper}

\end{document}